\documentclass[twocolumn,runningheads,natbib]{svjour2}
\smartqed                  
\usepackage{mathptmx}      
\usepackage{graphicx}

\newcommand{\be}{\begin{equation}}
\newcommand{\ee}{\end{equation}}
\newcommand{\ba}{\begin{eqnarray}}
\newcommand{\ea}{\end{eqnarray}}

\journalname{Astrophysics and Space Science}

\begin{document}

\title{Cooling of Neutron Stars with Strong Toroidal Magnetic Fields 
       \thanks{Work partially supported by UNAM-DGAPA grant \#IN119306}}

\titlerunning{Cooling of Neutron Stars with Strong Toroidal Magnetic Fields}

\author{Dany Page \and Ulrich Geppert \and Manfred K\"uker}

\institute{Dany Page \at
           Departamento de Astrof\'isica Te\'orica,
           Instituto de Astronom\'ia, 
           Universidad Nacional Aut\'onoma de M\'exico, 
           Mexico, D.F. 04510, Mexico \\
           \email{page@astroscu.unam.mx}
           \and
           Ulrich Geppert \at
           Departament de F\'isica Aplicada, 
           Universitat d'Alacant, 03080 Alacant, Spain \\
           \email{geppert@ua.es}
           \and
           Manfred K\"uker \at
           Astrophysikalisches Institut Potsdam, 
           An der Sternwarte 16, D-14482 Potsdam, Germany \\ 
           \email{mkueker@aip.de}
           }

\date{Received: 24 November 2006 / Accepted: 28 November 2006}

\maketitle

\begin{abstract}
We present models of temperature distribution in the crust of a neutron star
in the presence of a strong toroidal component superposed to the poloidal component
of the magnetic field.
The presence of such a toroidal field hinders heat flow toward the surface in a large
part of the crust.
As a result, the neutron star surface presents two warm regions surrounded by extended
cold regions and has a thermal luminosity much lower than in the case the magnetic field
is purely poloidal.
We apply these models to calculate the thermal evolution of such neutron stars
and show that the lowered photon luminosity naturally extends their life-time
as detectable thermal X-ray sources.

\keywords{Neutron star \and Magnetic field}
\PACS{97.60.Jd \and 26.60.+c \and 95.30.Tg}
\end{abstract}

\section{Introduction}
\label{intro}

It is generally considered that, within a few decades after its birth in a core
collapse supernova, a neutron star reaches a state of ``isothermality'' characterized
by a uniform high interior temperature.
The stellar surface is much colder and protected from the hot interior by a thin
($\sim$ 100 meters) layer, the envelope, which acts as a heat blanket.
As was argued by \cite{GH83}, in the presence of a sufficiently strong magnetic field, 
$> 10^{10}$ G, the surface temperature of the neutron star will not be uniform as
is expected in the unmagnetized case since the magnetic field severely limits the 
ability of electrons to transport heat in directions perpendicular to itself.
As a result, the regions around the magnetic poles, where the magnetic field is almost
radial, are expected to be significantly warmer than the regions around the magnetic
equator, where the field is almost tangent to the surface.
Since then much work has been dedicated to study the effects of the magnetic field on
the properties of the neutron star envelope and crust (see \citealt{PY01} and 
\citealt{PYCG03} for recent works).
In the presence of a sufficiently strong magnetic field, $\geq 10^{12} - 10^{13}$ G,
the anisotropy of heat transport actually extends to much higher densities 
and can even be present within the whole crust.
Recently, we have shown \citep{GKP04,GKP06} that, in cases where the field geometry in the 
crust is such that the meridional component of the field dominates over its radial 
component in a large part of the crust, the non-uniformity of the temperature, 
previously considered to be restricted to the envelope, may actually extend to the 
whole crust.
The largest effect was found when a strong toroidal component was included in
the crust, superposed to the poloidal component.
This result, that the geometry of the magnetic field {\em in the interior} of the neutron star
leaves an observable imprint at the surface, potentially allows us to study the internal
structure of the magnetic field through modelling of the spectra and pulse profiles of
thermally emitting neutron stars.

There exist growing observational evidence that the an\-isotropy of heat 
transport in the envelope alone, assuming an otherwise isothermal crust,
can not explain the surface temperature distributions of some observed neutron stars.
In the case of several of the ``Magnificent Seven" (see, e.g., \citealt{H04,H07})
optical broad band photometric detections can be interpreted as being due to the
Rayleigh-Jeans tail of a blackbody.
However, these optical data are well above the Rayleigh-Jeans tail of the
blackbody detected in the X-ray (``optical excess'') and indicate the presence of an
extended cold component of much larger area than the warm component observed in X-ray,
the latter having an emitting radius ($\sim 3-5$ km) much smaller than the usually assumed 
radius of a neutron star ($\sim 10 - 15$ km).
\cite{SHHM05} tried to fit the lightcurve of  RBS 1223 and concluded that
only a surface temperature profile with relatively small, about 4-5 kms across, 
hot polar regions may explain the observations. 
\cite{Petal02} and \cite{Tetal04} had arrived qualitatively at the same conclusion when they 
fitted the combined X-ray and optical spectrum of  RX J1856.5-3754.
In both cases, the smallness of the hot region is much below what can be reached
by considering anisotropic heat transport limited to only a thin envelope.

Little is known about the magnetic field structure in neutron stars which is very
likely determined by processes during the proto-neutron star phase and/or in a 
relatively short period after that epoch. 
A proto-neutron star dynamo \citep{TD93} is unlikely 
to generate purely poloidal fields while differential  rotation will easily
wrap any poloidal field and generate strong toroidal components 
\citep{KR98,WMW02}.
The magneto-rotational instability \citep{BH91} also most certainly  acts in 
proto-neutron stars \citep{AWML03} and results in toroidal fields from differential 
rotation \citep{BH98}.
Thus, it seems realistic to consider the effect of magnetic field 
configurations which consist of poloidal and toroidal components.

Besides their possible relevance for modelling the observed thermal radiation
of the ``Magnificent Seven'', and probably other strong field isolated
neutron stars, toroidal magnetic fields may have a strong effect on the thermal 
evolution of the stars. 
The highly non-uniform surface temperatures they can induce result in reduced
thermal luminosities and hence reduced energy losses during the late photon
cooling era.
It is our purpose in this work to explore this issue and we will present preliminary
models of cooling of neutron stars with strong toroidal fields.

The structure of this paper is the following.
In \S~\ref{sec:basic} we briefly review the basic ingredients and concepts involved
in modelling the cooling of a neutron star and in \S~\ref{sec:magfield} we present
in some detail the simple mathematical formalism which describes dipolar fields,
both poloidal and toroidal.
The next section, \S~\ref{sec:crust}, presents our results on the effect of strong magnetic
fields on the neutron star crust temperature distribution, and these results
are applied to model the cooling of the star in \S~\ref{sec:cooling}.
Finally, we discuss our results in \S~\ref{sec:conclusions} and offer some
tentative conclusions and prospects for future work.

\section{Basic mechanisms of neutron star cooling}
\label{sec:basic}

The essentials of neutron star cooling are expressed in the energy balance equation.
In its Newtonian formulation this equation reads
\be
\frac{dE_\mathrm{th}}{dt} = C_\mathrm{v} \frac{dT}{dt}
                   = -L_\nu - L_\gamma + H \, ,
\label{equ:energy-conservation}
\ee
where $E_\mathrm{th}$ is the thermal energy content of the star, 
$C_\mathrm{v}$ its total specific heat and $T$ its internal temperature, which
we assume to be uniform here.
The energy sinks are the neutrino luminosity $L_\nu$ and 
the surface photon luminosity $L_\gamma$.
The source term $H$ includes all possible ``heating mechanisms'' which will
be neglected in the present work;
however the decay of a strong magnetic field can result in significant heating
\citep{Kametal06,Kametal07}.
Only the basic points will be presented here and we refer the reader to the 
reviews \cite{YP04} and \cite{PGW06}, or \cite{PLPS04}, for more details.

The specific heat $C_\mathrm{v}$ receives its dominant contribution from the
baryons in the core of the star and about 10\% contribution from the leptons.
The crustal lattice, free neutrons in the inner crust and electrons also
provide a small contribution. 
When nucleons become superfluid (neutrons) or superconducting (protons) their
contribution to $C_\mathrm{v}$ is severely reduced and may even practically
vanish. 

The neutrino luminosity $L_\nu$ is usually dominated by the core.
We consider the slow emission processes of the modified Urca family and the
similar bremsstrahlung ones.
Once nucleon pairing turns on, neutrino emission from the formation and
breaking of Cooper pairs is also included.

\subsection{Neutron star envelopes (without magnetic fields)}
\label{sec:env}

Our main interest here is the photon luminosity $L_\gamma$.
In the absence of a magnetic field one expects the surface to have a uniform temperature
$T_\mathrm{e}$, the {\em effective temperature}, and can express $L_\gamma$ as
\be
L_\gamma = 4 \pi R^2 \, \sigma_\mathrm{SB} T_\mathrm{e}^4
\;\;\;\;\; \mathrm{or} \;\;\;\;\;
L_{\gamma \, \infty} = 4 \pi R_\infty^2 \, \sigma_\mathrm{SB} T_{\mathrm{e} \, \infty}^4
\label{Equ:l_gamma}
\ee
where $R$ is the star's radius and $\sigma_\mathrm{SB}$
the Stefan-Boltzmann constant
(the quantities ``at infinity'' are defined as
$L_{\gamma \, \infty} = e^{2\Phi} L_\gamma$,
$R_\infty = e^{-\Phi}\, R$, and 
$T_{\mathrm{e} \, \infty} = e^\Phi T_\mathrm{e}$, where
$e^\Phi \equiv (1-2GM/Rc^2)^{1/2}$ is the redshift factor).
In order to integrate Eq.~\ref{equ:energy-conservation} one needs a relationship
between $T$ and $T_\mathrm{e}$.
The assumption of uniform interior temperature $T$ is reasonable for most of the
interior, given the huge thermal conductivity of degenerate matter and once the
star is old enough to have relaxed from the initially complicated temperature
structure produced at its birth,  but is certainly not possible in the upper layers
of the star where density is low enough for the electrons not to be  fully degenerate.
One traditionally separates out these upper layers from the cooling calculation
and treats them as an {\em envelope}.
A typical cut density is $10^{10}$ g cm$^{-3}$ and the resulting envelope,
with a depth of the order of 100 meters, can be studied separately in a plane
parallel approximation.
\cite{GPE82} (and \citealt{GPE83}) presented detailed models of neutron star envelopes
and showed that $T_\mathrm{e}$ is related to the temperature at the bottom
of the envelope, $T_\mathrm{b}$ at density $\rho_\mathrm{b} = 10^{10}$ g cm$^{-3}$,
through the simple relation
\be
T_\mathrm{e} = 0.87 \times 10^6 \; \mathrm{K} \; 
               g_{s \, 14}^{1/4} \; T_\mathrm{b}^{0.55}
\label{equ:GPE}
\ee
where $g_{s \, 14}$ is the surface gravity in units of $10^{14}$ cm s$^{-2}$.
A relation such as Eq.~\ref{equ:GPE} is usually called a
``$T_\mathrm{b} - T_\mathrm{e}$ relationship''.
The models leading to Eq.~\ref{equ:GPE} assumed the envelope is formed of
iron-like nuclei, and it was shown by \cite{CPY97}
(see also \citealt{PCY97}) that if light elements, such as $H$, $He$, $C$, $O$, are
present deep enough in the envelope, the increase in thermal conductivity
(which is roughly proportional to $Z^{-1}$ in liquid matter, $Z$ being the
element's charge) results in much higher luminosities, by up to one order of
magnitude.
Since the temperature gradient penetrates deeper in hotter stars, larger 
amounts of light elements are necessary to alter heat transport at high
$T_\mathrm{b}$ than at lower $T_\mathrm{b}$.

\subsection{Heat transport with magnetic fields}
\label{sec:transport}

\begin{figure}
   \centering
   \includegraphics[width=0.40\textwidth]{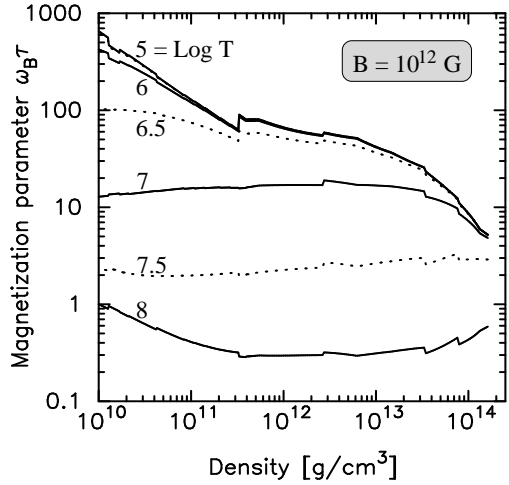}
   \caption{Magnetization parameter $\omega_{\scriptscriptstyle B} \tau$ 
            vs. density at six different temperatures (as labeled on
            the curves) assuming a uniform magnetic field 
            of strength $B =10^{12}$ G.
            Its value for different field strengths scales 
	    linearly in $B$.
            (Figure from \citealt{GKP04}.)}
   \label{fig:magnetization}
\end{figure}

In the absence of a magnetic field the thermal conductivity $\kappa$ is conveniently
written as
\be
\kappa_0 = \frac{1}{3} c_\mathrm{v} \overline{\mathrm{v}}^2 \tau
\;\;\;\;\; = \;\;\;\;\;
\frac{\pi^2 k_\mathrm{B}^2 T n_e}{3m_e^*} \tau
\ee
where $\overline{\mathrm{v}}$ is the mean velocity of the heat carriers,
$c_\mathrm{v}$ their specific heat per unit volume, and $\tau$ their collisional time;
the second expression is particularized to relativistic electrons, the dominant
heat carriers in neutron star crusts \citep{YU80}.
In the presence of a magnetic field, due to the classical Larmor rotation of electrons,
heat flow may be anisotropic and $\kappa$ becomes a tensor
\be
\kappa = \left(
         \begin{array}{ccc}
          \kappa_\perp  & \kappa_\wedge &    0      \\
         -\kappa_\wedge & \kappa_\perp  &    0      \\
                0       &       0       & \kappa_\| \\
         \end{array}
         \right)
\label{equ:kappa-tensor}
\ee
(assuming the field $\mathbf{B}$ oriented along the $z$-axis) whose
components have the form
\ba
\kappa_\|     &=& \kappa_0 \nonumber \\
\kappa_\perp  &=& \frac{\kappa_0}{1+(\omega_B \tau)^2} \label{equ:kappa2}\\
\kappa_\wedge &=& \frac{\kappa_0 \;\omega_B \tau}{1+(\omega_B \tau)^2} \nonumber
\ea
where $\omega_B = eB/m_e^* c$ is the electron cyclotron frequency.
The condition $\omega_B \tau \gg 1$, which implies strong anisotropy, is
easily realized in a neutron star envelope, and also possibly in the
whole crust \citep{GKP04}.
Values of the magnetization parameter $\omega_B \tau$ are plotted
in Fig.~\ref{fig:magnetization}.
Notice that $\tau$ receives contribution from both electron-phonon
and electron-impurity scattering, with frequencies $\nu_\mathrm{e-ph}$ and
$\nu_\mathrm{e-imp}$ resp., and is given by
$\tau = (\nu_\mathrm{e-ph}+ \nu_\mathrm{e-imp})^{-1}$.
Electron-impurity scattering dominates at low temperatures, and/or high densities,
and is $T$-independent while
$\nu_\mathrm{e-ph}$ goes roughly as $T^2$: combination of these two different
$T$-dependencies is the reason for the complex behavior of $\omega_\mathrm{B} \tau$
seen in Fig.~\ref{fig:magnetization}.

In case the field is strong enough to be quantizing, 
the expressions~\ref{equ:kappa2} have to be modified 
(in particular $\tau$ also becomes anisotropic) but the essential result
that $\kappa_\perp \ll \kappa_\|$, when $\omega_B \tau \gg 1$, remains \citep{P99}.

\begin{figure*}
   \centering
   \includegraphics[width=0.99\textwidth]{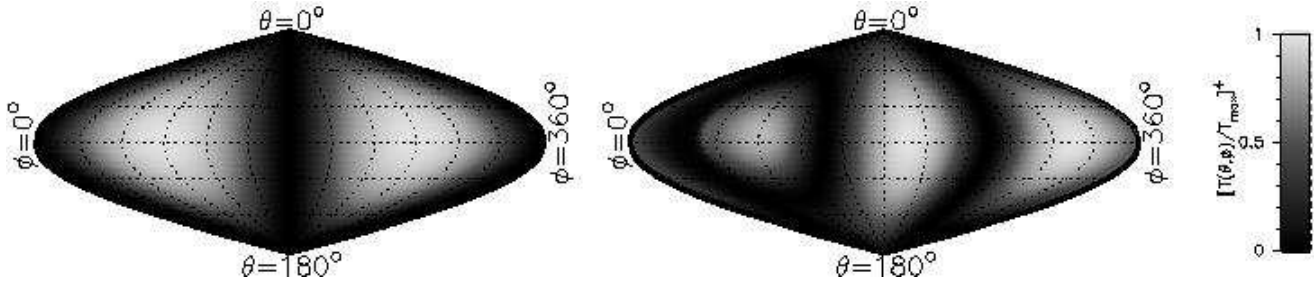}
\caption{Two examples of surface temperature distribution obtained from
         Eq.~\ref{Equ:GH}.
         The maps cover the whole neutron star surface in an area preserving
         projection.
         For better viewing, the magnetic field symmetry axis is located
         in the equatorial plane, oriented from $\phi = 90^\circ$ to
         $\phi = 270^\circ$.
         The left panel has a dipolar field only while the right panel
         also contains a quadrupolar field superposed to the same dipole.
         (Figure from \citealt{PGW06}.)
}
\label{Fig:Tdistr}
\end{figure*}

\subsection{Magnetized neutron star envelopes}
\label{sec:magenv}

In considering magnetic field effects on heat transport in a neutron star
the simplest case to handle is the envelope:
since its thickness, $\sim$ 100 m, is much smaller than the length scale over
which the field is expect to vary significantly, $\sim$ km, one can consider
heat transport on a given small patch on the surface to be independent of
the rest of the surface.
Moreover, the field strength $B$ and its angle with respect to the radial
direction, $\Theta_B$, can be considered as uniform in the patch and
the heat flux considered as essentially radial.
The thermal conductivity in a direction making an angle $\Theta_B$ with the field,
considering the $\kappa$ tensor of Eq.~\ref{equ:kappa-tensor}, is then given by
\be
\kappa(\Theta_B) = 
\cos^2 \Theta_B \times \kappa_\| + \sin^2 \Theta_B \times \kappa_\perp
\label{Eq:kappa_B}
\ee
With this form of $\kappa(\Theta_B)$ and a radial flux, heat transport in the
envelope at this surface patch is a one dimensional problem.
The solution is a ``$T_\mathrm{b} - T_\mathrm{e}$'' relationship which depends on
$B$ and $\Theta_B$, and the obtained effective temperature $T_\mathrm{e}$
is a local one which we will write as $T_\mathrm{s}(\theta,\phi)$ in the sense that the 
flux emerging from this patch is $\sigma_\mathrm{B} T^4_\mathrm{s}$.
Given the form of $\kappa(\Theta_B)$, Eq.~\ref{Eq:kappa_B},
\cite{GH83} proposed the simple interpolation formula
\ba
T_\mathrm{s}(\theta,\phi)^4 \equiv
T_\mathrm{s}(T_\mathrm{b};B,\Theta_B)^4  &\approx & \nonumber     \\
\cos^2 (\Theta_B) &\times& T_\mathrm{s}(T_\mathrm{b};B,\Theta_B=0)^4 + \nonumber \\
\sin^2 (\Theta_B) &\times& T_\mathrm{s}(T_\mathrm{b};B,\Theta_B=90^\circ)^4
\label{Equ:GH}
\ea
for arbitrary angle $\Theta_B$ in terms of the two cases of radial ($\Theta_B=0$) and tangential
($\Theta_B=90^\circ$) field.
Recently, \cite{PY01} and \cite{PYCG03}
have presented detailed calculations and fitted their results by an expression
similar to Eq.~\ref{Equ:GH}.

For a given geometry of the magnetic field (not necessarily dipolar) and
assuming that the temperature at the bottom of the envelope, $T_\mathrm{b}$,
is not affected by the magnetic field and hence is uniform arround the star,
one can generate the expected surface temperature distribution $T_s(\theta,\phi)$
by piecing together envelope models through applying relationships of the form of 
Eq.~\ref{Equ:GH} at each point of the surface (\citealt{P95,PS96}).
Two examples of such expected surface temperature distributions are shown in
Fig.~\ref{Fig:Tdistr}.
The photon luminosity is then given by
\be
L_\gamma = 4 \pi R^2 \cdot
\int \!\!\! \int \frac{d\Omega}{4\pi} \; \sigma_\mathrm{B} \; T_s(\theta,\phi)^4
\equiv 4 \pi R^2 \cdot \sigma_\mathrm{B} T_e^4
\label{equ:Teff}
\ee
where $T_\mathrm{e}$ is again defined as in Eq.~\ref{Equ:l_gamma}.

By this mehod one can easily generate surface temperature distributions 
corresponding to the geometry of the magnetic field at the surface (assuming
that the same field geometry is maintained throughout the underlying thin envelope).
However, the results of Fig.~\ref{fig:magnetization} show that the anisotropy
in heat transport is likely to extend much deeper into the crust than just
the envelope.
To handle such cases one is hence forced to define the field geometry in
the whole crust, which is what we do in the next section.

\section{The internal magnetic field}
\label{sec:magfield}

We will only consider here axisymmetric configurations
(mo\-re general ones will hopefully be considered in the future once a 3D transport
code is available).
In this case it is convenient to separate the magnetic field in two components
\be
\mathbf{B} = \mathbf{B}^\mathrm{pol} + \mathbf{B}^\mathrm{tor} \; ,
\ee
the poloidal and toroidal components, respectively,  
where $\mathbf{B}^\mathrm{pol}$ only has $\mathbf{e}_r$ and $\mathbf{e}_\theta$ 
components and $\mathbf{B}^\mathrm{tor}$ only an $\mathbf{e}_\phi$ 
component\footnote{Such decomposition is also possible for non-axisymmetric fields
but is more involved \protect\citep{R2000}.}
(the $\mathbf{e}_{r , \theta , \phi}$ 
being units vectors in the spherical coordinate directions,
with the $\theta = 0$ axis coinciding with the field symmetry axis).
Thus, the magnetic field lines of $\mathbf{B}^\mathrm{tor}$ are simply circles
centered on the symmetry axis but the field lines of $\mathbf{B}^\mathrm{pol}$
are more complicated.
Let us expand $B_r$ in Legendre polynomials as
$B_r = \sum \, F_l(r) \; P_l(\cos \theta)$.
We will only consider the first ($l=1$) term, the dipole, and write
$F_1$ as $S/r^2$ so that
\be
B_r = \frac{S(r)}{r^2} \cos \theta \; .
\label{equ:Stokes-r}
\ee
Then Maxwell's equation $\mathbf{\nabla \cdot B} = 0$ implies
\be
B_\theta = -\frac{1}{2r} \frac{\partial S}{\partial r} \sin \theta  \; .
\label{equ:Stokes-theta}
\ee
We also need the three boundary conditions
\begin{eqnarray}
S(r=0) &=& 0 
\nonumber
\\
S(r=R) = B_0 R^2
\;\;\; & \mathrm{and} & \;\;\;
\left. \frac{\partial S}{\partial r}\right|_{r=R} = -\frac{S(R)}{R} = -B_0 R
\end{eqnarray}
which ensure regularity at the star's center and smooth matching with
an external vacuum dipolar field, of strength $B_0$ at the magnetic poles,
at the stellar surface.
Besides these boundary conditions, the Stoke function $S$ is totally arbitrary
but a choice of it is equivalent to choosing the location of the currents sustaining
the poloidal field since the latter, having only a $j_\phi$ component, are given by
\be
j_\phi = \frac{c}{8\pi} \frac{\sin \theta}{r}
\left(\frac{\partial^2 S}{\partial r^2} - \frac{2S}{r^2}\right)
\ee
Notice also a simple physical interpretation of the Stoke function:
the magnetic flux through the star's equatorial plane in a circle of radius $r$
is simply given by $\pi S(r)$, and the boundary condition gives a total 
flux $\Phi = \pi R^2 B_0$, as it should be.
In vacuum, $S$ simply reduces to $B_0 R^3/r$.

\begin{figure}
\centering
  \includegraphics[width=0.35\textwidth]{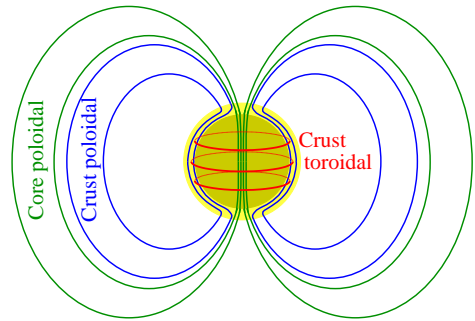}
\caption{The three components of the magnetic field considered in this work}
\label{Fig:fields}
\end{figure}

The locations of the currents in the stellar interior is unknown but it is
natural to separate them in two components, located in the core and in the crust
and accordingly separate $\mathbf{B}^\mathrm{pol}$ as
\be
\mathbf{B}^\mathrm{pol} = \mathbf{B}^\mathrm{core} + \mathbf{B}^\mathrm{crust} \; .
\ee

The core protons are expected to become a superconductor, with critical temperatures 
$T_c \sim 10^9$ K (see, e.g., \citealt{PLPS04}), soon after the star's birth:
the magnetic field is then confined into flux tubes and maintained by proton
supercurrents.
Since, by definition, $\mathbf{B}^\mathrm{core}$ corresponds to currents located
in the core, within the crust it is described by a vacuum dipolar poloidal field,
and we will parametrize it by its surface strength at the magnetic pole
$B_0^\mathrm{core}$.
We do not need to specify the distribution of the supercurrents sustaining
$\mathbf{B}^\mathrm{core}$ and the geometry of this component in the core since the
field, being confined to fluxoids which occupy only a very small volume, is not
expected to alter heat transport in the core.

For the $\mathbf{B}^\mathrm{crust}$ component, we need to specify its geometry in the
crust, i.e., the corresponding Stoke function $S^\mathrm{crust}$.
The arbitrariness involved in such specification can be somewhat relieved by 
considering models of the time evolution of this component:
currents spontaneously migrate toward the highest density region of the
crust, where the electrical resistivity is smallest, until they reach the
crust-core boundary where their migration is stopped by the proton superconductor
(see, e.g., Fig. 4 in \citealt{PGZ00}).
We use an $S^\mathrm{crust}$ function resulting from such evolutionary calculations
and scale it to vary the overall strength of $\mathbf{B}^\mathrm{crust}$,
which we parametrize by $B_0^\mathrm{crust}$ defined as its strength at the magnetic
pole.

\begin{figure}
   \centering
   \includegraphics[width=0.3\textwidth]{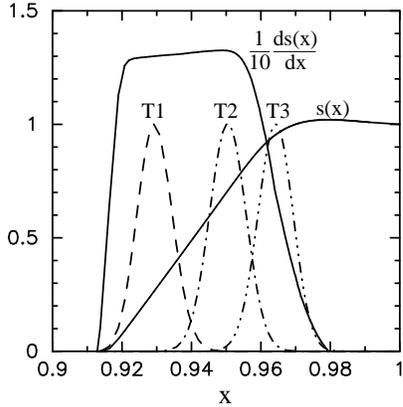}
   \caption{Continuous lines: normalized Stokes function 
    $s(x) = S/R^2B_0^\mathrm{crust}$ and its derivative
    (scaled by a factor 10) used in this work, which generate $B_r$ 
    (Eq.~\ref{equ:Stokes-r}) and $B_\theta$ (Eq.~\ref{equ:Stokes-theta}) resp.
    Discontinuous lines: the three different normalized functions 
    $t(x)=T/B_0^\mathrm{tor}$ we consider
    for the toroidal field, $B_\varphi$ (Eq.~\ref{equ:Stokes-tor}),
    which we label as T1, T2, and T3.
    On the horizontal axis $x \equiv r/R$.
    (Figure from \citealt{GKP06}.)
    }
 \label{fig:Stokes}
\end{figure}

For the toroidal component $\mathbf{B}^\mathrm{tor}$, we also expand it in Legendre
polinomials $P_l(\sin \theta)$ and keep only the $l=1$ dipolar term
\be
\mathbf{B}^\mathrm{tor} = T(r) \sin \theta
\label{equ:Stokes-tor}.
\ee
We do not have to consider the part of $\mathbf{B}^\mathrm{tor}$ confined to the
core, because of proton superconductivity, and specify only the part confined
to the crust.
The only restrictions are the boundary conditions
\be
T(r=R_\mathrm{core}) = 0
\;\;\;\;\; \mathrm{and} \;\;\;\;\;
T(r=R) = 0 \; .
\ee
We are not aware of evolutionary calculations of toroidal field in neutron star
crust and hence are left with a guess about the possible shape and size of
the $T$ function: 
following \cite{GKP06} we see it as a free parameter and consider several choices.
We parametrize the strength of $\mathbf{B}^\mathrm{tor}$, for each choice of $T$,
by $B_0^\mathrm{tor}$ defined as the maximum value of  $||\mathbf{B}^\mathrm{tor}||$
in the crust.

\cite{PAMP06} have presented similar models of field structure and the resulting
temperature distributions in the crust and chose force-free field configurations
to specify the field geometry.
It is reasonable to assume the field could evolve into a force-free configuration
during the early neutron star life, but there is no physical reason why the later
evolution, driven by Ohmic decay and Hall interactions, will conserve the force-free
condition and this motivates our considering the function $T$ as rather arbitrary.
However, our choice of $S^\mathrm{crust}$, inspite of being based on evolutionary
calculations of purely poloidal fields, may not be too realistic because of the
coupling between poloidal and toroidal components from the Hall term.
We show in Fig.~\ref{Fig:fields} a sketch of the field geometry we consider
and in Fig.~\ref{fig:Stokes} our choices of the functions $S$ and $T$.

\section{Magnetic field effects in the crust}
\label{sec:crust}

\begin{figure}
   \centering
   \includegraphics[width=0.30\textwidth]{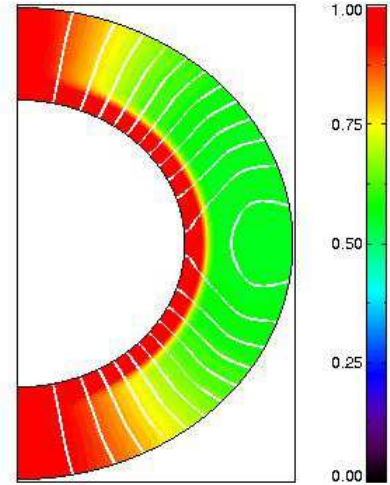}
   \caption{Temperature distribution in a strongly magnetized neutron star crust
   (whose thickness has been stretched by a factor five for easier reading).
    The choosen field scale parameters are
    $B_0^\mathrm{core}  = 7.5\times10^{13}$ G,
    $B_0^\mathrm{crust} = 2.5\times10^{13}$ G,
    $B_0^\mathrm{tor} =  3 \times10^{15}$ G,
    and the toroidal component's generating functions $T$ is the model "T1" of Fig.~\ref{fig:Stokes}).
    The color code maps the relative temperature, i.e., $T(r,\theta)/T_\mathrm{core}$,
    with a core temperature $T_\mathrm{core} = 6\times10^7$ K.
    White lines show field lines of $\mathbf{B}^\mathrm{pol}$, the field lines of 
    $\mathbf{B}^\mathrm{tor}$ being perpendicular to the plane of the figure.
    The heat blanketing effect of the toroidal component is clearly visible.
    (From \citealt{GKP06}.)}
   \label{fig:Tcrust}
\end{figure}

\begin{figure}
   \centering
   \includegraphics[width=0.18\textwidth,angle=-90]
       {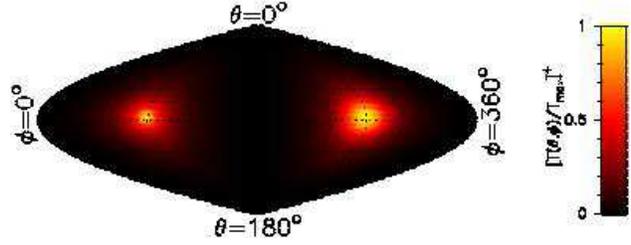}
   \caption{Surface temperature distribution corresponding to the crust temperature
    shown in Fig.~\ref{fig:Tcrust}.
    For better viewing the magnetic symmetry axis is in the equator 
    ($\theta = 90^\circ$) pointing at $\phi = 90^\circ$.
    (From \citealt{GKP06}.)}
   \label{fig:Tdist-surf}
\end{figure}

Given the magnetic field geometries described in the previous sections,
\cite{GKP06} calculated the resulting temperature distributions in the neutron 
star crust, in a stationary state.
In theses calculations, heat transport is solved in the crust for densities
from $\rho = \rho_\mathrm{core} = 1.6\times 10^{14}$ g cm$^{-3}$ down to
$\rho = \rho_\mathrm{b} \equiv 10^{10}$ g cm$^{-3}$.
In the core, at $\rho > \rho_\mathrm{core}$, the temperature $T_\mathrm{core}$
is assumed to be constant and uniform while at the outer zone, 
at $\rho = \rho_\mathrm{b}$, the models are matched with magnetized envelopes
(treated as the outer boundary condition).
Since the 2D transport code we use does not yet include correct specific heat 
it does not have the capability to perform realistc time dependent calculations and
only stationary results can be obtained, i.e., the thermal evolution is followed
until the temperature profile would not change anymore with time (this stationary
limit is independent of the specific heat).
Moreover, neutrino energy losses are also not included
(neutrino emission, for high temperature, and also strong magnetic fields,
is not negligible and should be included for accurate calculations, see, e.g.,
\citealt{PCY07}).
Fig.~\ref{fig:Tcrust} shows an example of the resulting temperature distribution
in the crust and Fig.~\ref{fig:Tdist-surf} shows the resulting surface temperature
distribution: this latter figure should be compared with the surface temperature distribution
shown in Fig.~\ref{Fig:Tdistr} where isothermal crusts were considered and magnetic
field affected heat transport only within the thin envelope.

\begin{figure*}
   \centering
   \includegraphics[width=0.90\textwidth]{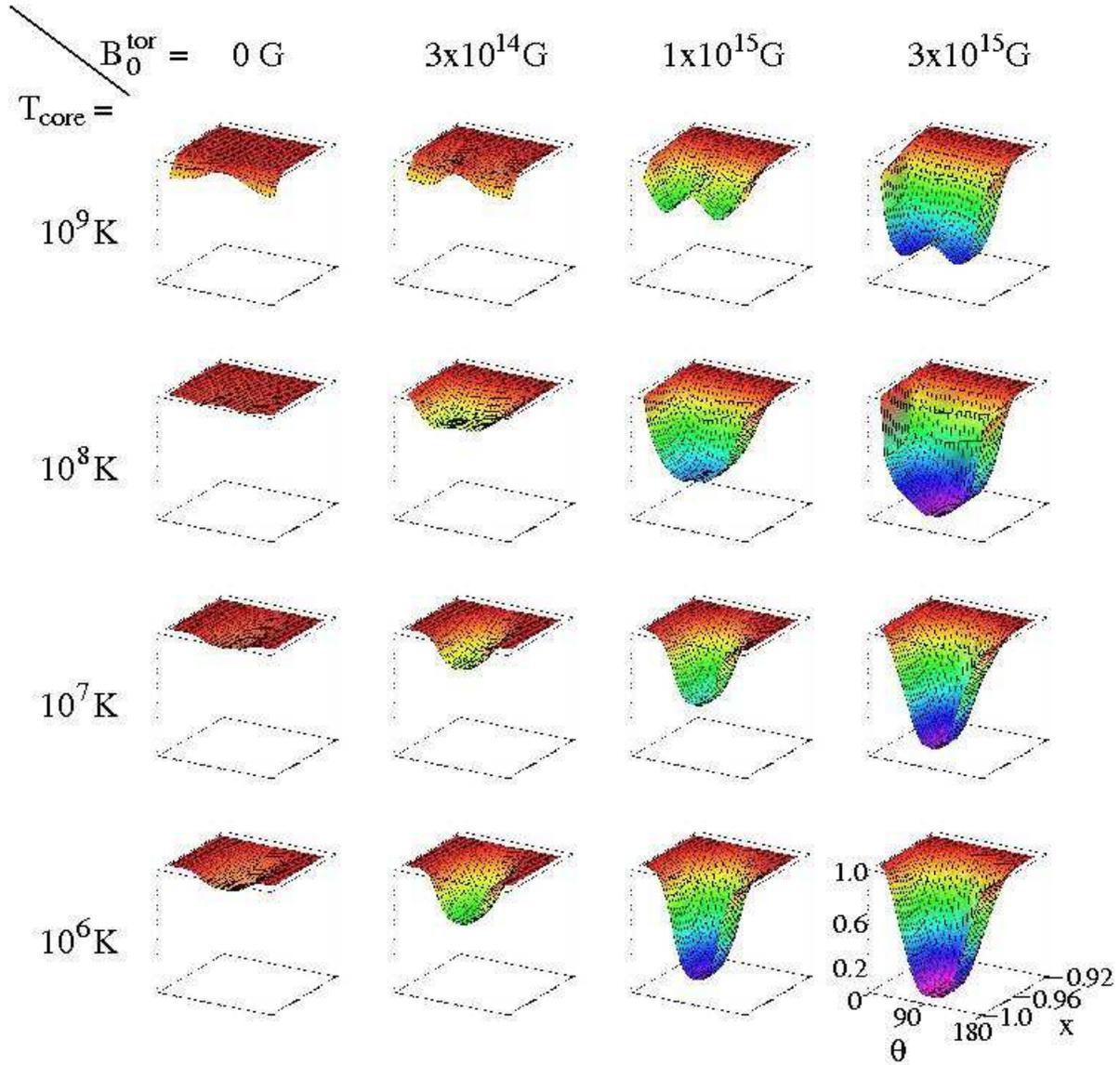}
   \caption{3D plots of the temperature distribution in a strongly magnetized 
    neutron star crust.
    The horizontal axes show radial coordinate $x=r/R$ and polar angle $\theta$
    while the vertical scale is $T/T_\mathrm{core}$.
    Four values of $T_\mathrm{core}$ are considered and
    the choosen field scale parameters for the poloidal field are
    $B_0^\mathrm{core}  = 2.5\times10^{13}$ G and
    $B_0^\mathrm{crust} = 7.5\times10^{13}$ G in all cases while the
    toroidal component scale $B_0^\mathrm{tor}$ is varied,
    the overall shape of the $T$ function, Eq.~\ref{equ:Stokes-tor},
    being the model ``T2'' of Fig.~\ref{fig:Stokes}.
    }
   \label{fig:Tcrust-all}
\end{figure*}

For a fixed magnetic field configuration and several values of $T_\mathrm{core}$, we
calculate the crustal temperature distribution and the resulting surface temperature
$T_\mathrm{s}(\theta)$, which is $\phi$ independent because of axial symmetry,
and obtain $T_\mathrm{e}$ from Eq.~\ref{equ:Teff}.
A set of results is displayed in Fig.~\ref{fig:Tcrust-all}:
we show models which, within our selection of field configurations, maximize
the magnetic field effects, with 75\% of the poloidal flux coming from crustal currents
and the toroidal component located in the middle of the crust.
With a purely poloidal field, i.e., $B_0^\mathrm{tor} = 0$, one sees little
effect and the crust stays close to isothermality, except at high temperature,
$T_\mathrm{core} = 10^9$ K, where the envelope's temperature gradient
extends deeper into the crust than the $\rho = 10^{10}$ g cm$^{-3}$ arbitrary cut.
Significant effects appear when $B_0^\mathrm{tor} \gg 10^{14}$ G and when
$B_0^\mathrm{tor} \geq 10^{15}$ G the crust is highly non-isothermal even at the
highest temperature $T_\mathrm{core} = 10^9$ since then 
$\omega_\mathrm{B} \tau \gg 1$ in the whole crust.
There is an intriguing change in the shape of the $T$-surface when going from
$T_\mathrm{core} > 10^8$ K to $T_\mathrm{core} < 10^8$ K, which is most evident
at the highest value $B_0^\mathrm{tor} = 3\times 10^{15}$ G:
it is most probably due to the shift in the dominant scattering process from
electron-phonon, at high $T$ where $\tau$ is $T$-dependent, to electron-impurity,
at low $T$ where $\tau$ is $T$-independent, as mentioned in Section~\ref{sec:crust}
(We hope to study this effect in more detail in a future work.).

\section{Cooling of strongly magnetized neutron stars}
\label{sec:cooling}

\begin{figure}
   \centering
   \includegraphics[width=0.45\textwidth]{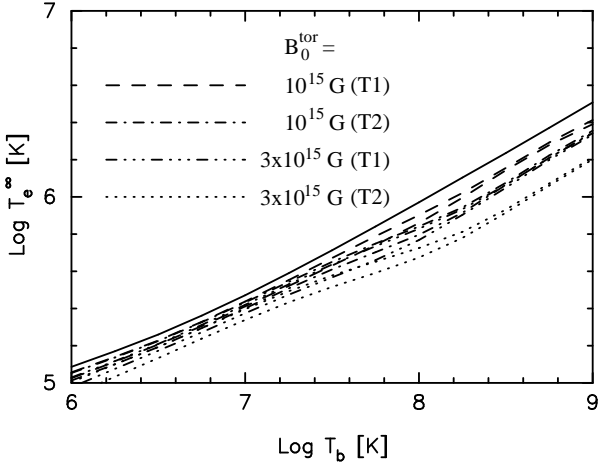}
   \caption{Possible $T_\mathrm{b} - T_\mathrm{e}$ relationships for a 
   neutron star with a strong toroidal magnetic field.
   The various sets of curves correspond to two different values of 
   $B_0^\mathrm{tor}$ and two different locations of the toroidal field,
   labelled as ``T1'' and ``T2'' as in Fig.~\ref{fig:Stokes}.
   For each case we also consider two different splitting of the poloidal field:
   $B_0^\mathrm{crust} = 7.5\times10^{12}$ G and 
   $B_0^\mathrm{core} = 2.5\times10^{12}$ G,
   or
   $B_0^\mathrm{crust} = 2.5\times10^{12}$ G and 
   $B_0^\mathrm{core} = 7.5\times10^{12}$ G,
   the cases with the larger $B_0^\mathrm{core}$ resulting in higher $T_\mathrm{e}$.
   The continuous cuve shows the $T_\mathrm{b} - T_\mathrm{e}$ relationships for an
   isothermal crust with the same poloidal field $B_0^\mathrm{pol} = 10^{13}$ G
   and magnetic field effects included only in the envelope \citep{PY01}.
   }
   \label{Fig:Tb-Te-crust}
\end{figure}

In order to perform (time dependent) cooling calculations we will here use the
models described in the previous section to produce a set of outer boundary conditions
for our 1D cooling code (see, e.g., \citealt{PLPS04}).
We fix the outer boundary at a density $\rho_\mathrm{b} = 10^{14}$ g cm$^{-3}$,
instead of $10^{10}$ g cm$^{-3}$ when envelope models are used as boundary conditions.
As is obvious from Fig.~\ref{fig:Tdist-surf}, compared to Fig.~\ref{Fig:Tdistr},
the photon luminosity, and hence $T_\mathrm{e}$, is much lower in presence of a
strong toroidal field than when the crust is considered as isothermal and
Fig.~\ref{Fig:Tb-Te-crust} shows the resulting $T_\mathrm{b}-T_\mathrm{e}$
relationships obtained for nine different magnetic field configurations.
For the most extreme case, $T_\mathrm{e}$ can be reduced by a factor of 2.5, and
hence $L_\gamma$ by a factor 40, compared to the isothermal crust case.
Notice that we chose $\rho_\mathrm{b} = 10^{14}$ g cm$^{-3}$ inspite of our crustal
models starting at $\rho_\mathrm{core} = 1.6\times 10^{14}$ g cm$^{-3}$ because we
found a very small temperature gradient in the range $1.0 - 1.6\times 10^{14}$ g cm$^{-3}$
and hence prefer to treat this density range with the cooling code.
The 1D cooling code solves the energy balance and heat transport equations in their
full general relativistic forms.

The cooling models we will consider here are within the minimal cooling paradigm of
\cite{PLPS04}.
In short, this means neutrino emission in the stellar core is from the modified Urca
and the similar nucleon bremsstrahlung processes, neutron and proton pairing is taken into
account with the resulting neutrino emission from the Cooper pair breaking and formation
process and the alteration of the specific heat.
We use pairing critical tempeatures for proton $^1S_0$ from \cite{T73},
for neutron $^1S_0$ from \cite{SFB03}, and for neutron $^3P_2$ the 
model ``a'' from \cite{PLPS04}.
The star model is a 1.4 M$_\odot$ neutron star built with the equation of state
of \cite{APR98} and hence, as part of the minimal cooling paradigm, charged
meson condensates\footnote{The APR equation of state shows the presence of a $\pi^0$
condensate, but it has no noticeable effect on the cooling.},
hyperons, and deconfined quark matter are not present in the star.

\begin{figure}[t]
   \centering
   \includegraphics[width=0.44\textwidth]{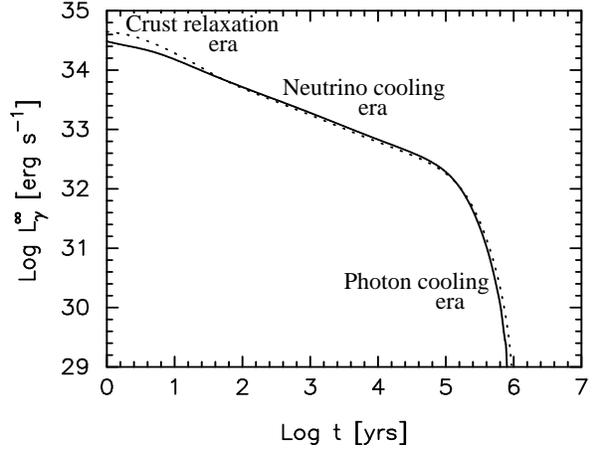}
   \caption{Comparison of cooling trajectories,
   i.e., red-shifted surface photon luminosity vs time,
   obtained with our truncated models
   using $\rho_\mathrm{b} = 10^{14}$ g cm$^{-3}$ (continuous curve) vs. a ``full'' model
   having $\rho_\mathrm{b} = 10^{10}$ g cm$^{-3}$ (dotted curve).
   Both models consider a poloidal dipolar field of strength $B_0^\mathrm{pol} = 10^{13}$ G.
   The truncated model uses the $T_\mathrm{b} - T_\mathrm{e}$ relationship for an isothermal
   crust shown in Fig.~\ref{Fig:Tb-Te-crust} while the full model uses directly the
   $T_\mathrm{b} - T_\mathrm{e}$ relationship from \cite{PY01}.
   }
   \label{Fig:Cool_test}
\end{figure}

First we assess the effect of the approximation of truncating the star at 
$\rho_\mathrm{b} = 10^{14}$ g cm$^{-3}$, and treating most of the crust through
our stationary solutions as part of the outer boundary condition, 
instead of using the traditional outer boundary at $10^{10}$ g cm$^{-3}$.
A comparison of cooling trajectories with these two different boundary densities,
and in the absence of a toroidal field component,
is shown in Fig.~\ref{Fig:Cool_test}.
The major difference appears at early times during the crust relaxation era:
a ``full'' (i.e., using  $\rho_\mathrm{b} = 10^{10}$ g cm$^{-3}$)
model shows strong radial temperature radients in the curst at this stage
(see, e.g., \citealt{GYP01}) while the truncated 
(i.e., using  $\rho_\mathrm{b} = 10^{14}$ g cm$^{-3}$)
model assumes an isothermal crust.
Another way of seing it is that our 2D crustal models only considered stationary states
and are hence only applicable when the cooling time scale is much longer than the
thermal relaxation time of the crust, a condition which is certainly not fulfilled
during this early cooling phase.
Later, during the neutrino cooling era, the truncated model is slightly warmer
than the ``full'' model because neutrino emission from its truncated crust is
missing.
During the photon cooling era the difference between the two models becomes larger,
now due to the smaller specific heat of the truncated model which consequently cools
faster.
Overall, moving the outer boundary from $10^{10}$ g cm$^{-3}$ to $10^{14}$ g cm$^{-3}$
only has a small, and quite negligible, effect except at early ages which we will hence
not show in our next results.

\begin{figure}[t]
   \centering
   \includegraphics[width=0.42\textwidth]{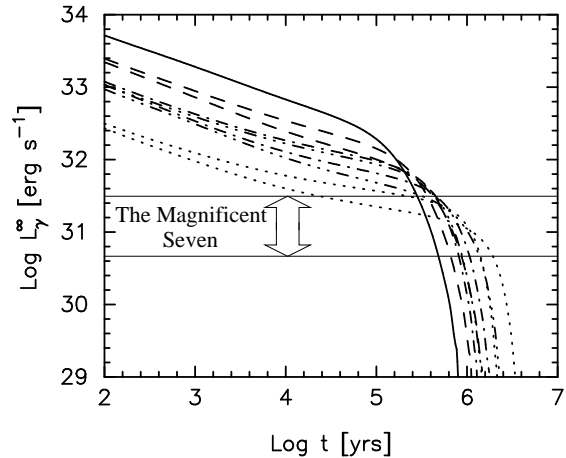}
   \caption{Cooling of neutron stars with strong toroidal magnetic fields:
   red-shifted surface photon luminosity vs time.
   Each curve corresponds to one of the $T_\mathrm{b}-T_\mathrm{e}$ relationships
   displayed in Fig.~\ref{Fig:Tb-Te-crust}, with the same line-style.
   }
   \label{Fig:Cool_final}
\end{figure}

Having now the confidence that stripping the star of most of its crust, and properly including
the stripped part into the outer boundary condition, introduces an acceptably small error,
we can proceed with models having strong toroidal crustal fields.
We show in Fig.~\ref{Fig:Cool_final} our results for the nine field configurations 
of the $T_\mathrm{b}-T_\mathrm{e}$ relationships presented in the Fig.~\ref{Fig:Tb-Te-crust}.
During the neutrino cooling era, the star's core evolution is driven by 
$L_\nu$ and the surface temperature, and $L_\gamma$, simply follows the evolution of the
core.
So during this phase all nine models have exactly the same evolution but they look very 
different at the surface: models with a higher $T_\mathrm{e}$ for a given $T_\mathrm{b}$,
as shown in Fig.~\ref{Fig:Tb-Te-crust}, have a higher photon luminosity. 
During the photon cooling era the results are inverted since $L_\gamma$ drives the
cooling and models with a higher $T_\mathrm{e}$ for a given $T_\mathrm{b}$ result in 
a larger $L_\gamma$ and consequently they undergo faster cooling.

We have preferred to plot the cooling curves as $L_\gamma$ vs $t$ instead of $T_\mathrm{e}$ vs $t$:
given the highly non-uniform $T_\mathrm{s}(\theta,\phi)$ the effective temperature $T_\mathrm{e}$
looses any observational meaning.
More detailed models taking this into account will be presented in a future work \citep{PHG07}.

\section{Discussion and Conclusions}
\label{sec:conclusions}

We have extended our previous results about temperature distribution in a strongly
magnetized neutron star crust \citep{GKP04,GKP06} to a broader range of temperatures
and applied them to study the cooling of neutron stars with strong toroidal magnetic
fields.
\cite{GKP04} had shown that with a purely poloidal field entirely confined to the
crust strongly non-uniform temperature distributions develop in the crust. Then 
\cite{GKP06} showed that allowing part of this poloidal field to permeate the core
significantly reduced the temperature non-uniformity but that the inclusion of a strong
toroidal field component can result in crustal temperature gradients even stronger
than in the first case (similar results have been obtained by \citealt{PAMP06}).
We have shown here that, due to the strength of the field, the magnetization parameter
$\omega_{\scriptscriptstyle B} \tau$ is very large in most of the crust even at 
core temperatures as high as $10^9$ K and the large crustal temperature gradients
are still present in such hot stars.
Considering these results we have been able to perform cooling calculations of
such neutron stars with a strong toroidal crustal magnetic field, with some approximations
which, as we have shown, introduce only very small errors in the results.

Our final results are displayed in Fig.~\ref{Fig:Cool_final} and compared with the
estimated luminosity range of the ``Magnificent Sev\-en'' \citep{H04}.
This comparison indicates most probable ages between 0.5 and 1.5 Myrs for these stars,
depending on the magnetic field structure, but this range could be extended down to
0.05 Myrs for the brightest ones and up to 3 Myrs for the dimest ones if the most
extreme field geometry is considered.
Besides the uncertainty in the field structure, which in itself results in the range
of predicted ages shown in Fig.~\ref{Fig:Cool_final}, there are two other key ingredients
in cooling models which can have similar effects:
nucleon pairing in the core and the chemical composition of the envelope.
The former significantly affects the specific heat during the photon cooling era
and different assumptions about the values of $T_\mathrm{c}$ can introduce a factor of
a few in the predicted cooling ages while the latter can significantly affect the photon
luminosity (see discussion in \S~\ref{sec:env}) and introduce another uncertainty of
a few (we refer the reader to the detailled presentation of \citealt{PLPS04}).
Considering these three theoretical uncertainties, magnetic field geometry, nucleon 
pairing, and envelope chemical composition, it is certainly possible to extend the
theoretical cooling ages, for photon luminosities between $\sim 3\times 10^{30}$
and $\sim 3\times 10^{31}$, to cover the range from $10^5$ to almost $10^7$ years !

Can we reasonably expect to reduce this enormous theoretical age uncertainty?
Precession of some neutron stars \citep{L07,P07} may imply that neutron pairing
critical temperatures in the core are very low \citep{L03}, 
a possibility which moreover has a strong theoretical fundament \citep{SF04}: 
this would imply a large specific heat and favor long cooling times (within the minimal
cooling paradigm).
Interpretation of the absorption lines detected in most of the ``Magnificent Seven'' 
(see, e.g., \citealt{H07} and \citealt{vKK07}) may provide crucial information about
the surface chemical composition.
However, atomic and molecular physics in strong field still has many secrets to
be unveiled \citep{T07} and, moreover, what is needed is the chemical composition of 
the deep\-er layers of the envelope, i.e., several tens of meters below the surface.
Finally, the structure and evolution of the magnetic field with toroidal components,
in the neutron star context as considered in this paper, is an almost uncharted territory
and much progress can be expected.

\begin{acknowledgements}
We thank Jos\'e Pons and many other participants of the meeting ``Isolated Neutron Stars''
for discussions,
as well as Jillian A. Henderson for careful and critical reading of this paper.
\end{acknowledgements}



\end{document}